\documentclass[3p,times]{elsarticle}

\usepackage{ecrc}

\usepackage{xspace}
\usepackage[margin=0.5ex,caption=false]{subfig}

\usepackage{epstopdf}

\volume{00}

\firstpage{1}

\journalname{Nuclear Physics A}

\runauth{Benjamin Bannier (for the PHENIX Collaboration)}

\jid{nupha}

\jnltitlelogo{Nuclear Physics A}

\usepackage{graphicx}
\usepackage{amsmath,amssymb}

\usepackage{lineno}

\newcommand{\pt}{\ensuremath{{p_T}}\xspace}
\newcommand{\Rg}{\ensuremath{R_\gamma}\xspace}
\newcommand{\ef}{\ensuremath{\langle \varepsilon f\rangle}\xspace}
\newcommand{\piz}{\ensuremath{{\pi^{0}}}\xspace}
\newcommand{\Npart}{\ensuremath{N_\text{part}}\xspace}

\begin{document}

\begin{frontmatter}



  \title{Systematic studies of the centrality dependence of soft photon production in Au+Au collision with PHENIX}

  \author{Benjamin Bannier (for the PHENIX\fnref{col1} Collaboration)}
  \fntext[col1]{A list of members of the PHENIX Collaboration and acknowledgments can be found at the end of this issue.}
  \address{Department of Physics and Astronomy, Stony Brook University, Stony Brook, NY 11794, USA}




  \begin{abstract}
    Since the earliest days of Heavy Ion Physics thermal soft photon radiation
    emitted during the reaction had been theorized as a smoking gun signal for
    formation of a quark-gluon plasma and as a tool to characterize its properties.
    In recent years the existence of excess photon radiation in heavy ion
    collisions over the expectation from initial hard interactions has been
    confirmed at both RHIC and LHC energies by PHENIX and ALICE respectively. There
    the radiation has been found to exhibit elliptic flow $v_2$ well above what can
    currently be reconciled with a picture of early emission from a plasma phase.
    During the 2007 and 2010 Au+Au runs PHENIX has measured a high purity sample of
    soft photons down to $\pt>0.4\,\text{GeV}/c$ using an external conversion
    method. We present recent systematic studies by PHENIX from that sample on the
    centrality dependence of the soft photon yield, and elliptic and triangular
    flow $v_2$ and $v_3$ in Au+Au collisions which fill in the experimental picture
    and enable discrimination of competing soft photon production scenarios.
  \end{abstract}

  \begin{keyword}
  \end{keyword}

\end{frontmatter}



\section{Introduction}
\label{sec:intro}

Collisions of heavy ions create dense and hot states of hadronic medium like
the Quark Gluon Plasma (QGP), for which direct photons, i.e.\ photons not
produced in decays of hadrons, provide an excellent probe: they can be produced
during all stages of the interaction, and leave the medium virtually
undisturbed due their vanishing interaction cross section with the hadronic
medium.
They can thus probe the full space-time evolution of the system.
Here photons with transverse momenta $\pt = \sqrt{p_x^2 + p_y^2} \gtrsim
2.5\,\text{GeV}/c$ are produced predominately in early, hard interactions and
called \emph{hard} photons, while photons with smaller transverse momentum are
called \emph{soft} and thought to be produced from the medium.
The photon yield gives experimental access to the rates of their different
production processes, and their correlation with the event geometry, i.e.\
their elliptical and triangular flow coefficients $v_2$ and $v_3$, are
sensitive to the dynamics of the medium. While direct photon yield and flow
integrate over all production channels, both quantities measured together
strongly constrain possible production scenarios.

\begin{figure}
  \subfloat[%
    \label{fig:atm}
    Correlation between pair mass assuming production of electron
    and positron at the nominal interaction point $M_\text{cgl}$, and at the
    HBD shell $M_\text{atm}$. The concentration of photons with
    $10\,\text{MeV}/c^2<M_\text{cgl}<15\,\text{MeV}/c^2$ and
    $M_\text{atm}<5\,\text{MeV}/c^2$ corresponds to conversions in the HBD
    shell; the opposite concentration at
    $10\,\text{MeV}/c^2<M_\text{atm}<15\,\text{MeV}/c^2$ and
    $M_\text{cgl}<5\,\text{MeV}/c^2$ is due to decay photons from \piz Dalitz
  decays, $\piz\rightarrow\gamma(ee)$.]{
  \includegraphics[width=.5\linewidth, clip, trim={.6cm .6cm 0 0}]{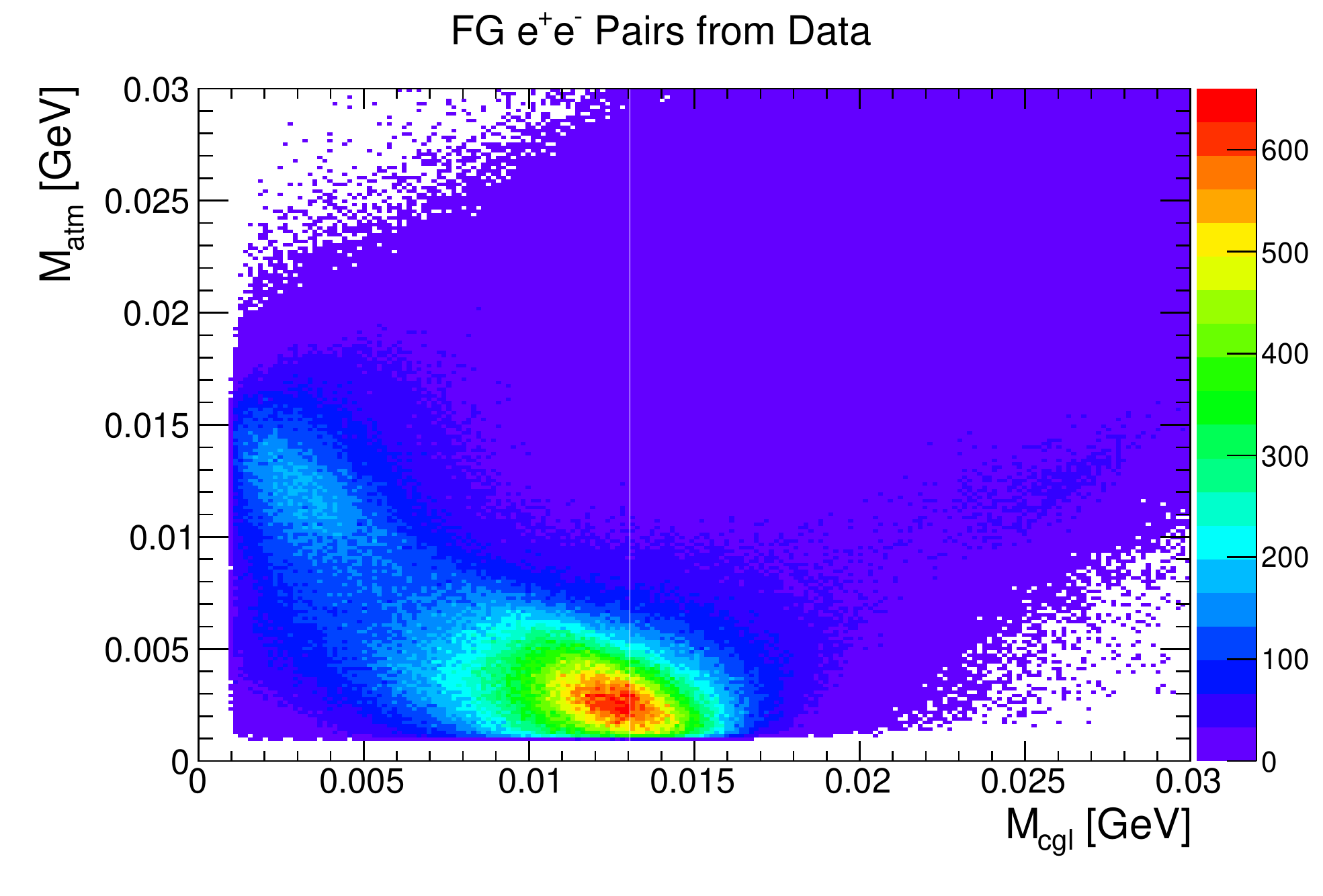}
  }
  \subfloat[%
    \label{fig:rg}
    \Rg for different centrality classes from the 2007 and 2010 runs,
    compared to the virtual photon result from PHENIX \cite{PPG086}.
    Here and later statistical uncertainties are shown as bars, systematic
    uncertainties as boxes.  Both results agree within uncertainties.]{
    \includegraphics[width=0.5\linewidth, clip, trim={.8cm 0 1.7cm 1.3cm}]{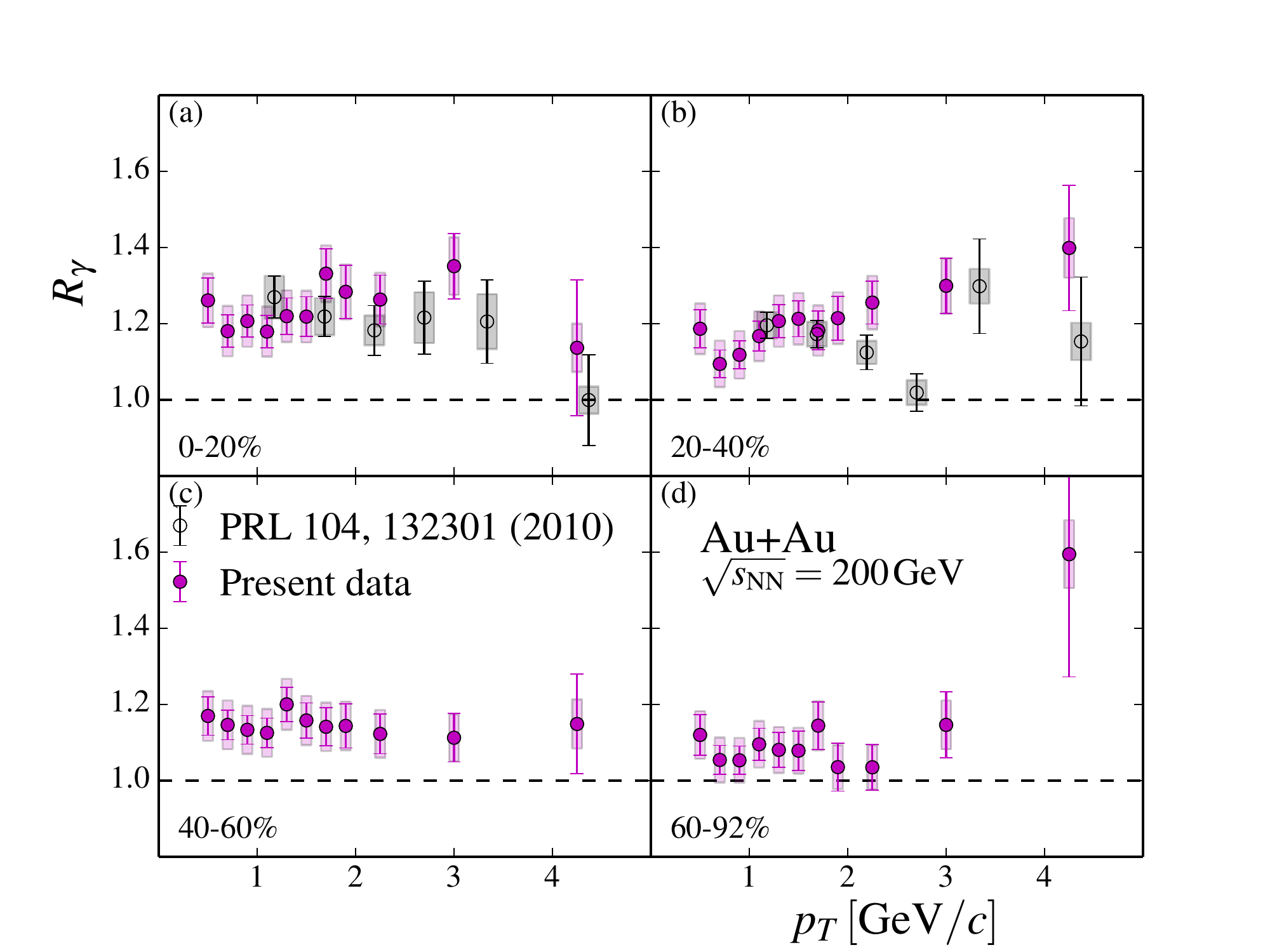}
  }
  \caption{}
\end{figure}

Measurements of soft photons are notoriously difficult in electromagnetic
calorimeters due to large contamination from misidentified hadrons and a
deteriorating energy resolution.
Instead we here reconstruct photons from external conversions to electrons and
positrons.
For the 2007 and 2010 RHIC runs the HBD detector was installed in the PHENIX
detector providing spatially well-defined conversion locations on a cylindrical
shell $R=60\,\text{cm}$ from the beam pipe with $X/X_0=2$ to $3\%$
\cite{Anderson201135}.
We identify conversion pairs by their characteristic apparent pair masses
(opening angles) at the nominal interaction vertex and at the HBD detector
shell.  Momenta of electrons and positrons can be calculated assuming they came
from either the vertex or the HBD shell, and their invariant pair mass (opening
angle) at the vertex and the HBD shell can be compared, see Fig.~\ref{fig:atm}.
Applying a simultaneous selection on both mass variables allows a clean
separation of electron-positron pairs from Dalitz decays and external photon
conversions with the level of background $<1\%$ in the conversion sample while
maintaining a good photon momentum resolution.

\section{Direct photon yield}
\label{sec:yield}

While by selecting on the two apparent pair masses we have already arrived at a
very pure photon sample its relation to the actual photon production
rate $Y_\gamma^\text{incl}$ depends on detector-specific quantities, namely the
photon conversion probability $p_\text{conv}$, the geometrical acceptance of
the detector for electron-positron pairs $a_{ee}$, and the efficiency of the
used experimental cuts $\varepsilon_{ee}$, with the exact values only
known inside relatively large systematic uncertainties.
Instead of attempting to accurately determine these correction factors we
follow a different approach: in addition to the raw yield of photons
$N_\gamma^{\text{incl}}$, we measure a raw yield of photons from the
decay $\piz\rightarrow\gamma\gamma$, $N_\gamma^\piz$ which we extract by
pairing one photon reconstructed in a conversion pair with another photon
reconstructed in the PHENIX calorimeters with very loose cuts and estimation of
the combinatorial background with mixed-event photon-photon pairs. The raw
\piz-tagged yield $N_\gamma^\piz$ is related to the yield
$Y_\gamma^\piz$ by the same detector-dependent factors as the inclusive photon
yield, and additionally a conditional acceptance factor \ef quantifying the
probability to reconstruct both photons from a \piz decay, given that one
photon was already reconstructed in a conversion pair. Since we use only very
loose cuts to select calorimeter photons we can trade less dependence on
systematic uncertainties for reduced statistical significance via the
signal-to-background ratio in the \piz-tagged sample.
The shared factors then drop out in the ratio of both quantities so that we can
formulate a quantity \Rg,
\begin{equation} \Rg = \frac{Y_\gamma^\text{incl}}{Y_\gamma^\text{decay}} =
  \frac{Y_\gamma^\text{incl}/Y_\gamma^\piz}
  {Y_\gamma^\text{decay}/Y_\gamma^\piz} =
  \frac{\frac{N_\gamma^\text{incl}/{p_\text{conv}a_{ee}\varepsilon_{ee}}}{N_\gamma^\piz/{p_\text{conv}a_{ee}\varepsilon_{ee}}\ef}}{\frac{Y_\gamma^\text{decay}}{Y_\gamma^{\pi^0}}}
  = \frac{\ef
  \frac{N_\gamma^\text{incl}}{N_\gamma^{\pi^0}}}{\frac{Y_\gamma^\text{decay}}{Y_\gamma^{\pi^0}}}
\label{eq:rg} \end{equation}
Here we have used the yield of photons from the decay of any
hadron $Y_\gamma^\text{decay}$ which can be calculated from the known hadron
yields and their branching ratios to photons. The numerator of the RHS of
Eq.~(\ref{eq:rg}) depends only on measured raw yields and the conditional
acceptance \ef which has to be determined in a Monte Carlo simulation of the
detector; the denominator depends on known yields and branching ratios and can
be calculate in e.g.\ a simple phase space simulation. With these definition
any measurement $\Rg>1$ corresponds to a direct photon signal. Our results for
\Rg are show in Fig.~\ref{fig:rg}. We observe a substantial direct photon
signal.

\begin{figure}
  \subfloat[%
    \label{fig:yield_spectra}
    Direct photon \pt spectra for different centrality classes. The shaded
    bands indicate $N_\text{coll}$-scaled fits to PHENIX $pp$ data. While the
    direct photon yield varies over two orders of magnitude between
    centralities the shape shows little change, also see the text.]{
    \includegraphics[width=0.5\linewidth]{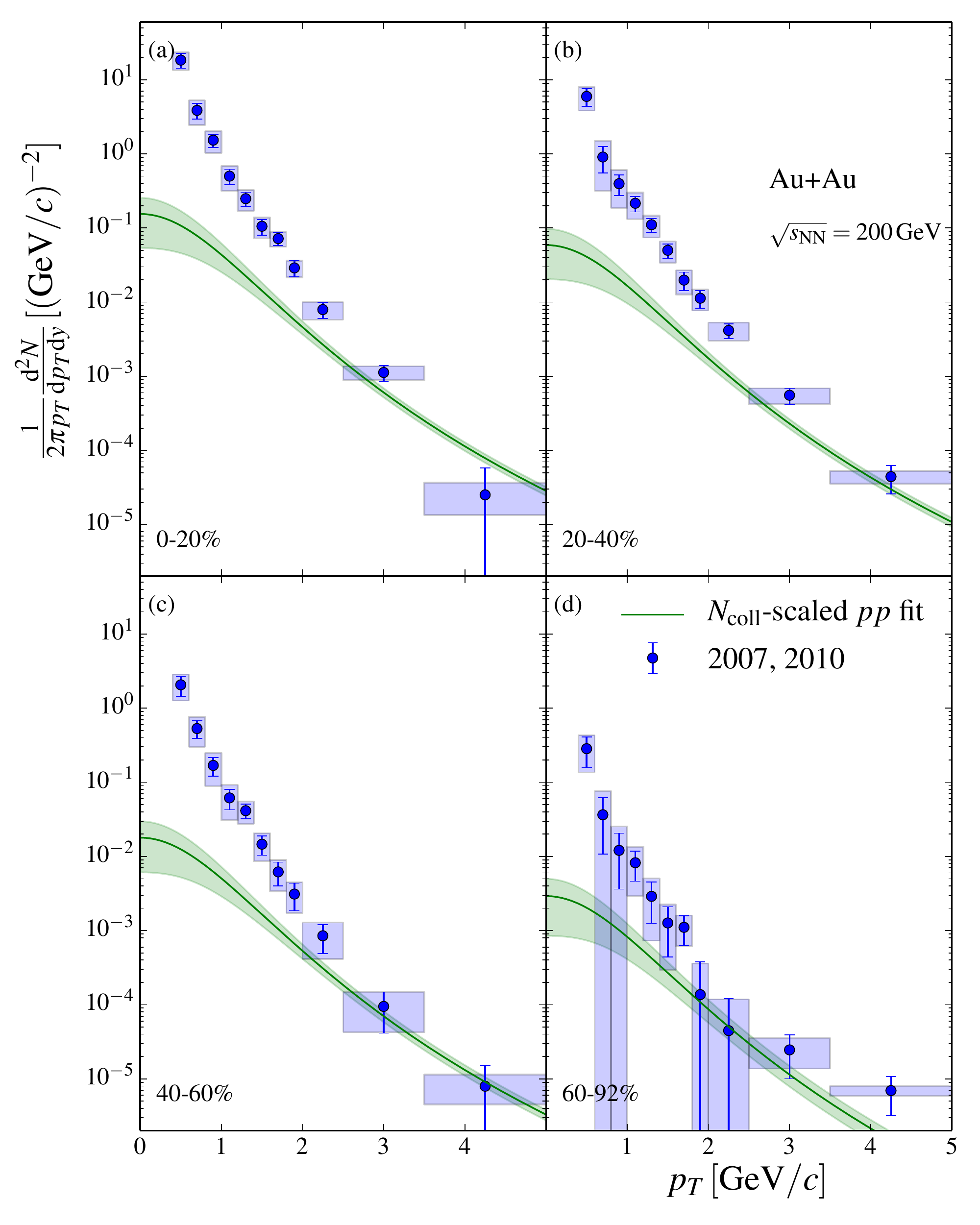}
  }
  \subfloat[%
    \label{fig:yield_int}
    The \pt-integrated direct photon yield for different lower integration
    limits, note logarithmic axes. The dashed lines indicate independent fits
    to each set up measurements. The centrality-dependence of the integrated
    direct photon yield shows no dependence on the lower integration limit
    outside of uncertainties.]{
    \includegraphics[width=0.5\linewidth]{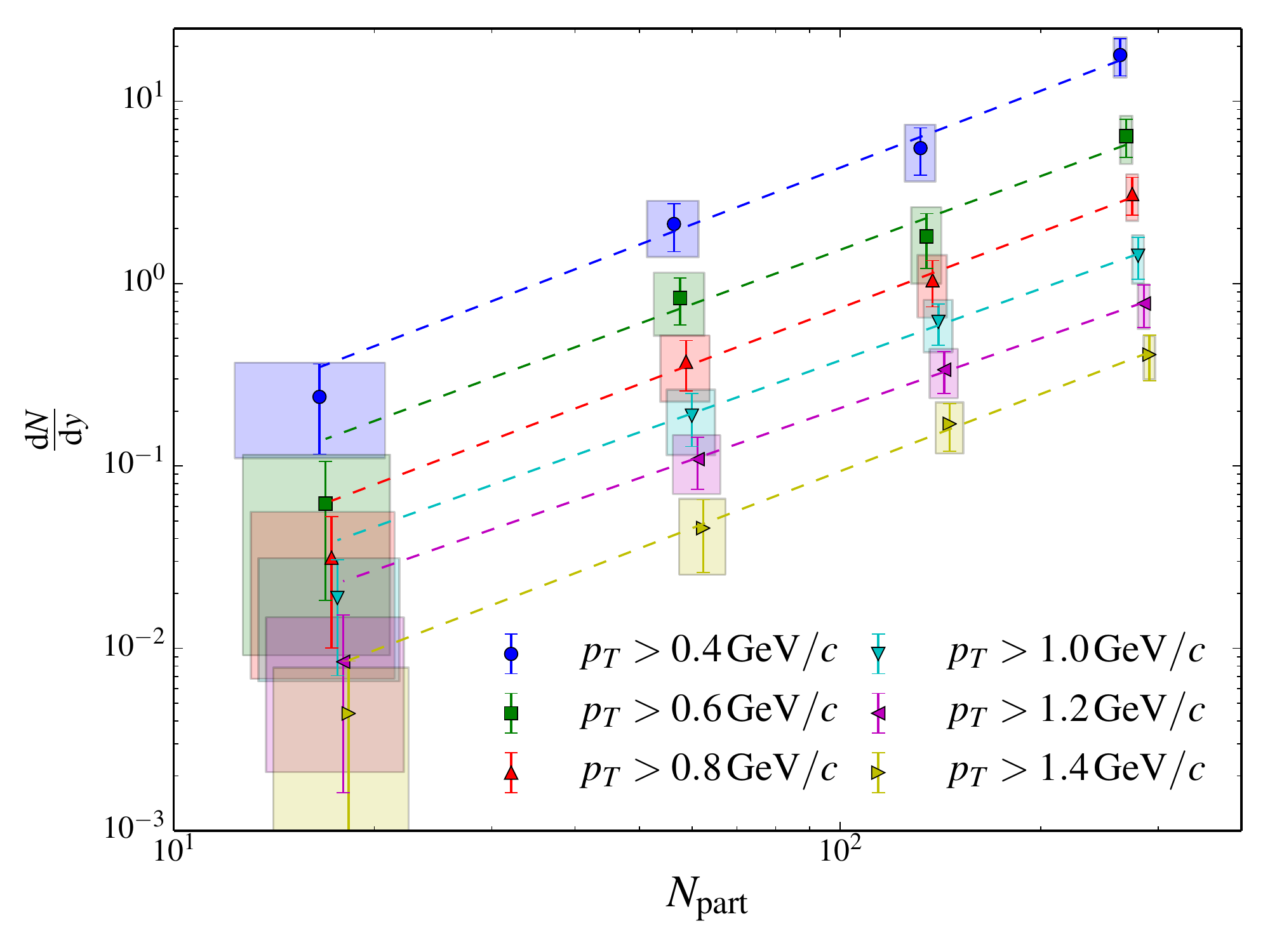}
  }
\caption{Direct photon yield}
\label{fig:yield}
\end{figure}

From \Rg we can calculate the direct photon yield shown in
Fig.~\ref{fig:yield_spectra},
\begin{equation} Y_\gamma^\text{direct} = (\Rg - 1) Y_\gamma^\text{decay}
\end{equation}
and analyze its centrality-dependence. We find that the direct photon excess
over the $N_\text{coll}$-scaled $pp$ yield has inverse slopes roughly
independent of centrality, $(239\pm 25\pm 7)\,\text{MeV}/c$ (0-20\%), $(260\pm
33\pm 8)\,\text{MeV}/c$ (20-40\%), $(225\pm 28\pm 6)\,\text{MeV}/c$ (40-60\%),
and $(238\pm 50\pm 6)\,\text{MeV}/c$ (60-92\%), and the \pt-integrated yield of
the excess over the $N_\text{coll}$-scaled $pp$ yield has a power-law
dependence on the number of participants \Npart, $N_\gamma\propto
\Npart^\alpha$ with a power larger than that of hadrons, $\alpha={1.48\pm
0.08(\text{stat})\pm 0.04(\text{syst})}$, see Fig.~\ref{fig:yield_int}.

\section{Direct photon $v_2$ and $v_3$}
\label{sec:vn}

\begin{figure}
  \centering
  \includegraphics[width=0.7\linewidth]{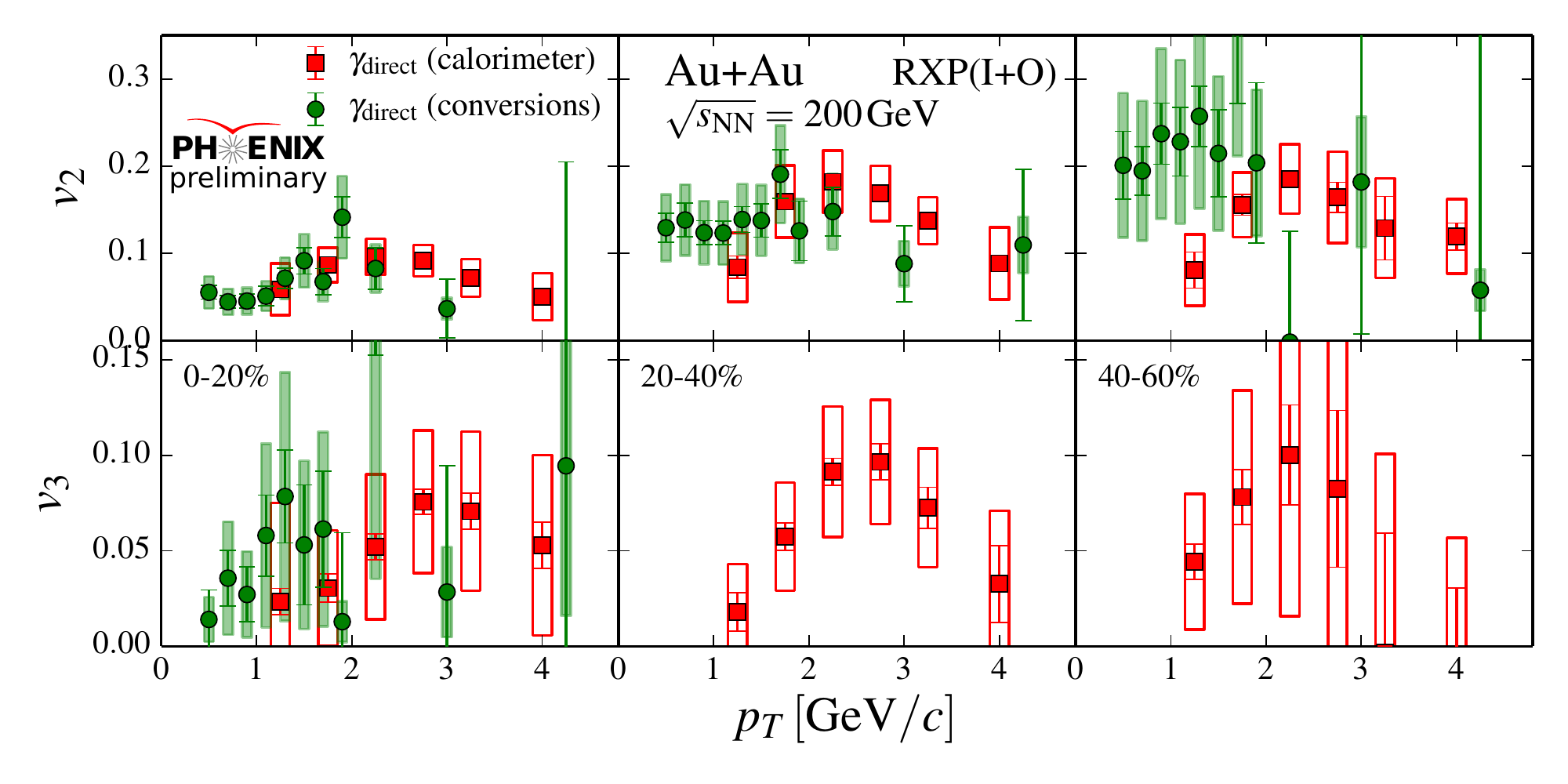}
  \caption{%
    The direct photon $v_2$ (\emph{top}) and $v_3$ (\emph{bottom}) as a
    function of the photon \pt in different centrality classes. The results
    from this analysis are shown with circle markers, from a preliminary
    calorimeter analysis as squares.
    Both results are consistent within systematic uncertainties.
  }
  \label{fig:vn}
\end{figure}

The elliptical and triangular flow coefficients $v_2$ and $v_3$ of direct
photons can be calculated from the raw inclusive photon flow coefficients
$v_n^\text{incl}$, the expectation for photons from hadron decays
$v_n^\text{decay}$, and the known composition of the inclusive photon sample
quantified by \Rg,
\begin{equation} v_n^\text{direct} = \frac{\Rg v_n^\text{incl} -
v_n^\text{decay}}{\Rg -1} \end{equation}
Here the $v_n$ can be calculated from the angles between the $n$-th order event
plane measured at forward rapidities $1.0<|\eta|<2.8$, $\psi_n$, and the photon
direction $\phi_n$ with a Fourier decomposition, $v_n^\prime = \langle\cos
2(\phi_n - \psi_n)\rangle$.  To obtain the actual $v_n$ we perform resolution
corrections of the raw $v_n^\prime$ with the 3-subevent method
\cite{Poskanzer:1998yz}, taking the difference to results from a 2-subevent
resolution correction into account the systematic uncertainties.
The $v_n^\text{decay}$ can be calculated in a phase space simulation from the
known $v_n$ and yields of the parent hadrons and their branching ratios to
photons by measuring the decay photon $\phi_n$ against the known event planes,
i.e.\ with perfect resolution. Our result for the direct photon $v_2$ and $v_3$
are shown in Fig.~\ref{fig:vn}. We observe markedly positive, non-zero
coefficients which remain large down to low \pt across all centralities.

\section{Conclusions}
\label{sec:conclusions}

We have extracted a high-purity sample of soft photons and simultaneously
measured the direct photon yield and the direct photon elliptical and
triangular flow coefficients $v_2$ and $v_3$, and extended the measurements in
the soft regime.
We find a substantial direct photon signal consistent with an earlier
measurement using virtual photons \cite{PPG086}, and with \pt-integrated yields
growing with the number of participants \Npart faster than the yield of soft
hadrons.  The shape of the direct photon spectra shows no changes outside of
uncertainties across centralities.
The coefficients of the elliptical flow measured in the same direct photon
sample show markedly positive values, consistent with results from a virtual
photon analysis \cite{PPG126} and preliminary results from an analysis of
photons measured in the PHENIX calorimeters. While as one might expected from
the more eccentric collision geometry we find increasing $v_2$ values when
going to more peripheral collision, the $v_2$ of direct photons appears to show
less \pt dependence than that of soft hadrons, even with an indication of
flattening towards the smallest \pt as already indicated by earlier
measurements~\cite{Drees:2013wza}.
%








%
%
%

\bibliographystyle{apsrev}
\bibliography{bib}

\end{document}